# Design and performance of LLRF system for CSNS/RCS*

LI Xiao1)   SUN Hong   LONG Wei
ZHAO Fa-Cheng   ZHANG Chun-Lin
Institute of High Energy Physics, Chinese Academy of Sciences, Beijing 100049, China

**Abstract**: The rapid cycling synchrotron (RCS) is part of China Spallation Neutron Source (CSNS). The RCS provides 1.6GeV protons with a repetition rate of 25Hz. The RF system in RCS is mainly composed of a ferrite loaded RF cavity, a high power tetrode amplifier, a bias supply of 3300A and a digital low level RF (LLRF) system based on FPGA. The major challenge of the LLRF system is to solve problems caused by rapid frequency sweeping and heavy beam loading effect. This paper will present the design and structure of the LLRF system, and show results of performance tests.

## 1 Introduction

China Spallation Neutron Source (CSNS)[1,2] is composed of an H- linac and a proton rapid cycling synchrotron (RCS). It is designed to accelerate proton beam pluses to 1.6 GeV, striking a metal target to produce spallation neutron for scientific research. The injection and extraction beam energy in CSNS/RCS are 80MeV and 1.6GeV respectively. Table 1 summarizes the primary parameters related to RF system.

Table 1. RF system parameters for CSNS/RCS.

| Parameters | Value |
|---|---|
| Beam power (kW) | 100 |
| Circumference (m) | 229.7 |
| Energy (GeV) | 0.081~1.6 |
| Intensity (ppp) | 1.87E13 |
| Circulating dc current (A) | 1.5/ 3.6 |
| Repetition rate (Hz) | 25 |
| Harmonic number | 2 |
| RF frequency (MHz) | 1.02~2.44 |
| Peak RF voltage (kV) | 165 (h=2) |

A total of 8 fundamental RF systems are used in CSNS/RCS to provide a peak RF voltage of 165kV. The RF systems operate on the harmonic number h=2, and the frequency sweeps from 1.022MHz at injection to 2.444MHz at extraction with a repetition rate of 25Hz. Fig.1 shows the frequency, amplitude and synchronous phase patterns during an accelerating time of 20ms. According to the working pattern, the frequency, the amplitude and the phase of the RF signal should be carefully controlled. The CSNS/RCS LLRF control system is designed to achieve required acceleration voltage amplitude and phase regulation of ±1% and ±1degree respectively. The resonant state of the cavity and the high beam loading effect should also be taken care of by LLRF system, to make sure of the stability of the RF system.

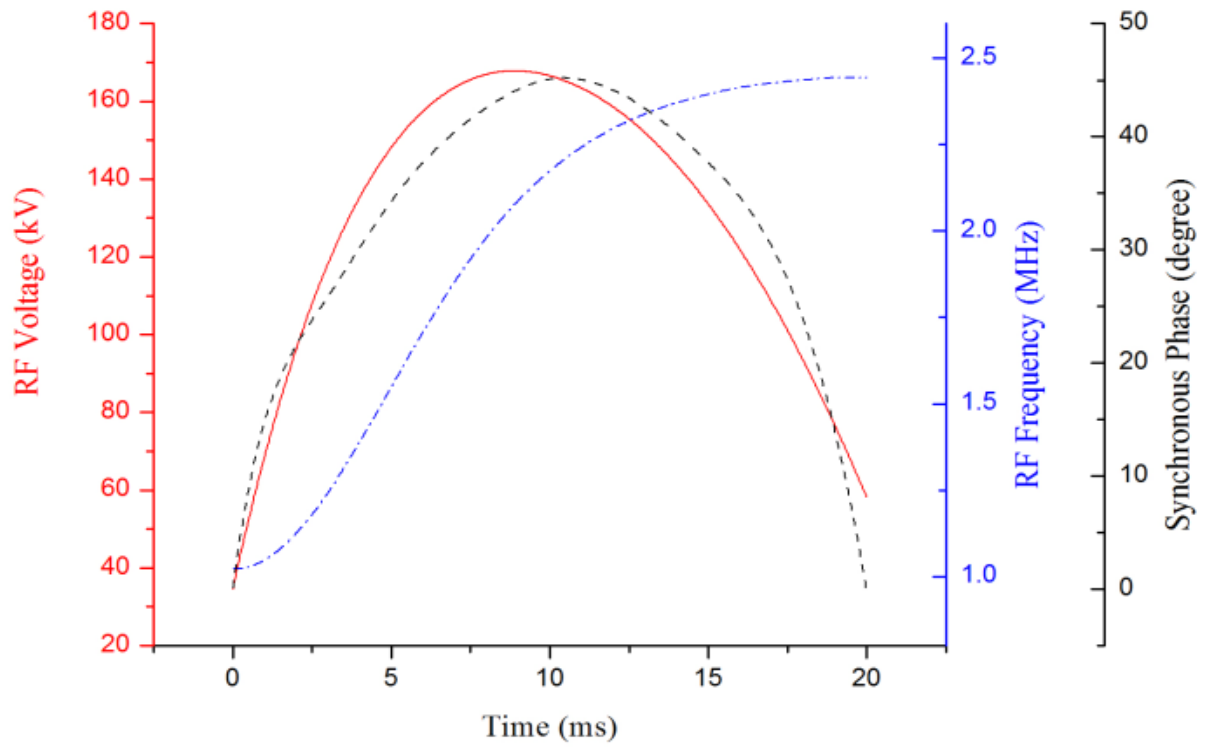

Fig. 1. RF frequency (dot-dash line), accelerating voltage (solid line) and synchronous phase (dot line) in one cycle.

## 2 CSNS/RCS RF system

In CSNS/RCS, each RF system is mainly composed of a cavity, a bias supply, a high power RF amplifier and a digital LLRF system[3]. The cavity used is a ferrite-loaded coaxial resonator with 2 accelerating gaps and single ended. The loaded material is Ferroxcube 4M2. The inductance of cavity can be shifted as a bias field altering. Corresponding to the bias current varying range of 200~3000A, the resonant frequency of the cavity can sweep from 1.02 to 2.44MHz to satisfy the CSNS/RCS operation. For one gap, a nominal peak RF gap voltage of more than 11.8kV is required. In order to satisfy the requirement of cavity dynamic tuning caused by the nonlinear characteristics of the ferrite material during rapid sweeping of RF frequency and voltage, a 3300A switch type power supply and a 150A linear type power supply are used in bias supply. Two types power supply are connected in parallel, and a big range more than 3000A and a high bandwidth up to 20 kHz can be achieved respectively. The RF amplifier consists of a three stage amplifiers chain. The final stage amplifier using a tetrode TH558 operated in class AB1, with a configuration of cathode-grounded. The maximum plate dissipation of the tube is 500kW. The tetrode is driven by a feedback (FB) amplifier of 800W located in the rack of final stage amplifier adjacent to the cavity. The RF drive signal from a wide-band solid-state preamplifier (SSA) of 500W will be combined with a RF signal extracted from the cavity which having a proper amplitude and delay to drive FB amplifier. Fig.3 shows the simplified schematic of CSNS/RCS RF system.

* Supported by National Natural Science Foundation of China (11175194)
1) Email: lixiao@ihep.ac.cn

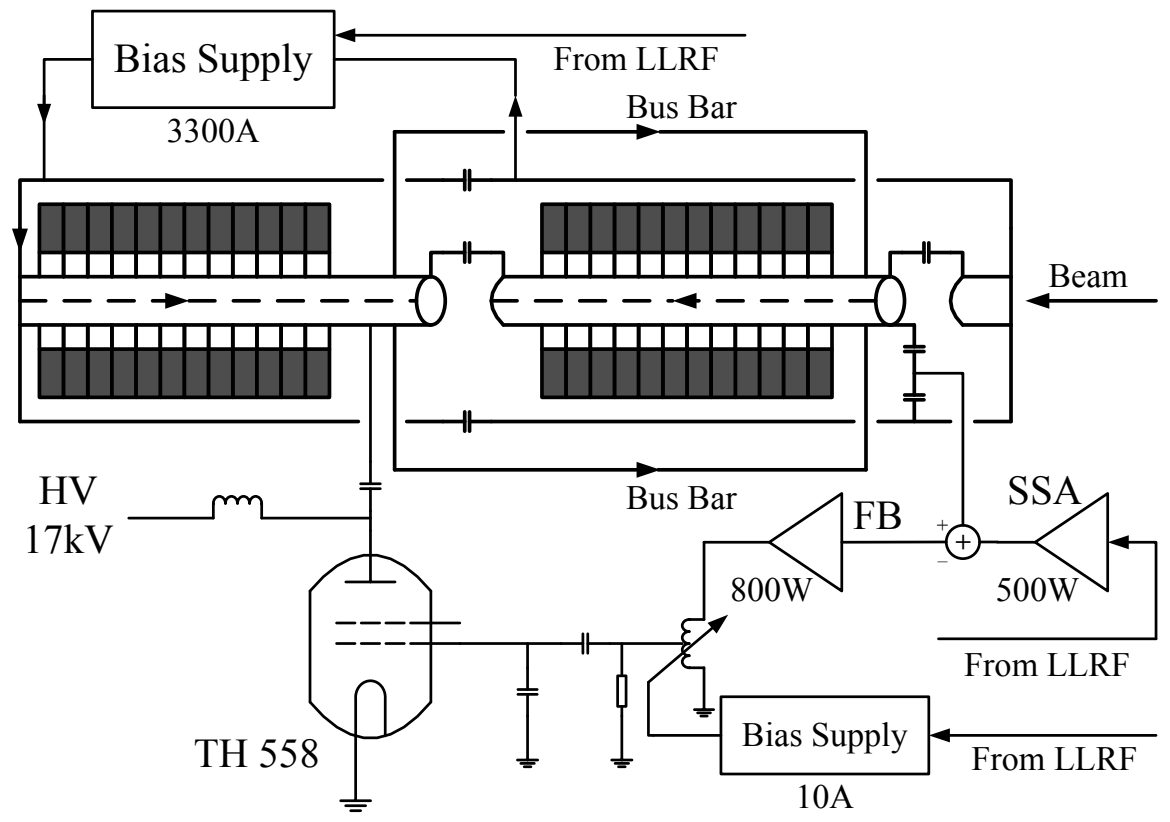

Fig. 2. CSNS/RCS RF system simplified schematic.

## 3 LLRF system overview[4]

Considering system compatibility and bandwidth, CPCI bus is adopted. The hardware platform includes the CPCI6200 CPU, the CSNS standard timing board to receive system clock and event trigger, the custom CPCI carrier to realize RF signal digitization, data processing and control arithmetic for control function. The CPCI hardware is arranged into two CPCI creates. One is for acceleration voltage control and cavity dynamic tuning. Another is for beam orbit correction and beam loading compensation.

The custom CPCI carrier is the heart of LLRF control system. It features one Stratix IV EP4SGX530 FPGA which provides 531.2K equivalent logic elements. It also includes high performance TMS320C6655 DSP, PCI9656 CPCI bridge chip, high speed ADC and DAC with sampling rate up to 125MHz, optical interface, ethernet interface, and so on. Operation command and control data exchange protocol between the CPCI6200 and the CPCI carriers were carefully defined. The custom driver was developed to access specified registers on CPCI carriers from CPCI6200 running the VxWorks operating system. Adhering to the CSNS control standard, the LLRF control system is Experimental Physics and Industrial Control System (EPICS)-based. EPICS drivers have been developed to provide user interface through standard Channel Access protocol. As shown in Fig.3, each custom CPCI carrier has an allocated time slot of 2.25ms for transmission data in 25Hz system operation and the actual bandwidth is greater than 200MB/s. A self-defined bus based on LVDS is adopted to transmit control data, which is shared by the CPCI carriers. Event trigger and following clock are used to ensure the reliability and stability of transmission, and the maximum bandwidth is greater than 3.84 Gb/s. Fig. 4 shows the configuration of the LLRF control system.

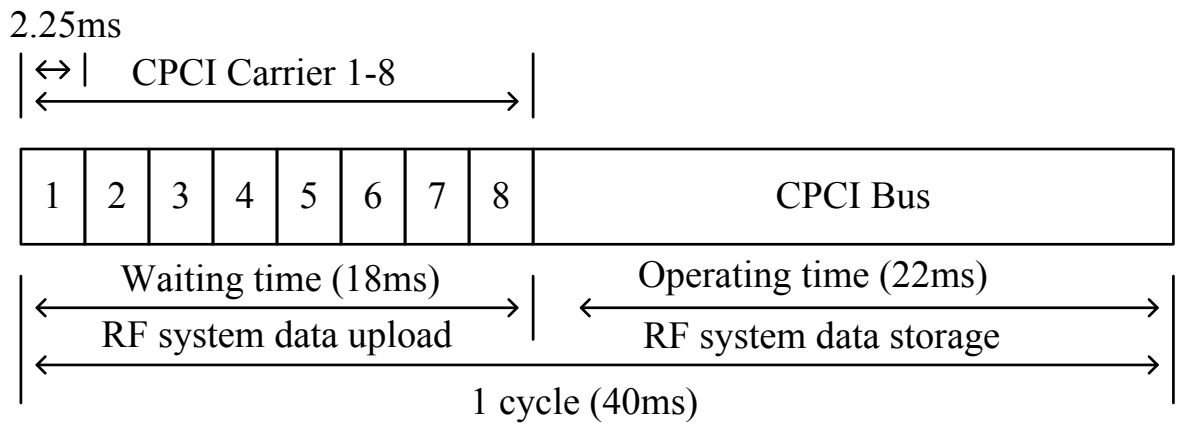

Fig. 3. CPCI bus operation timing

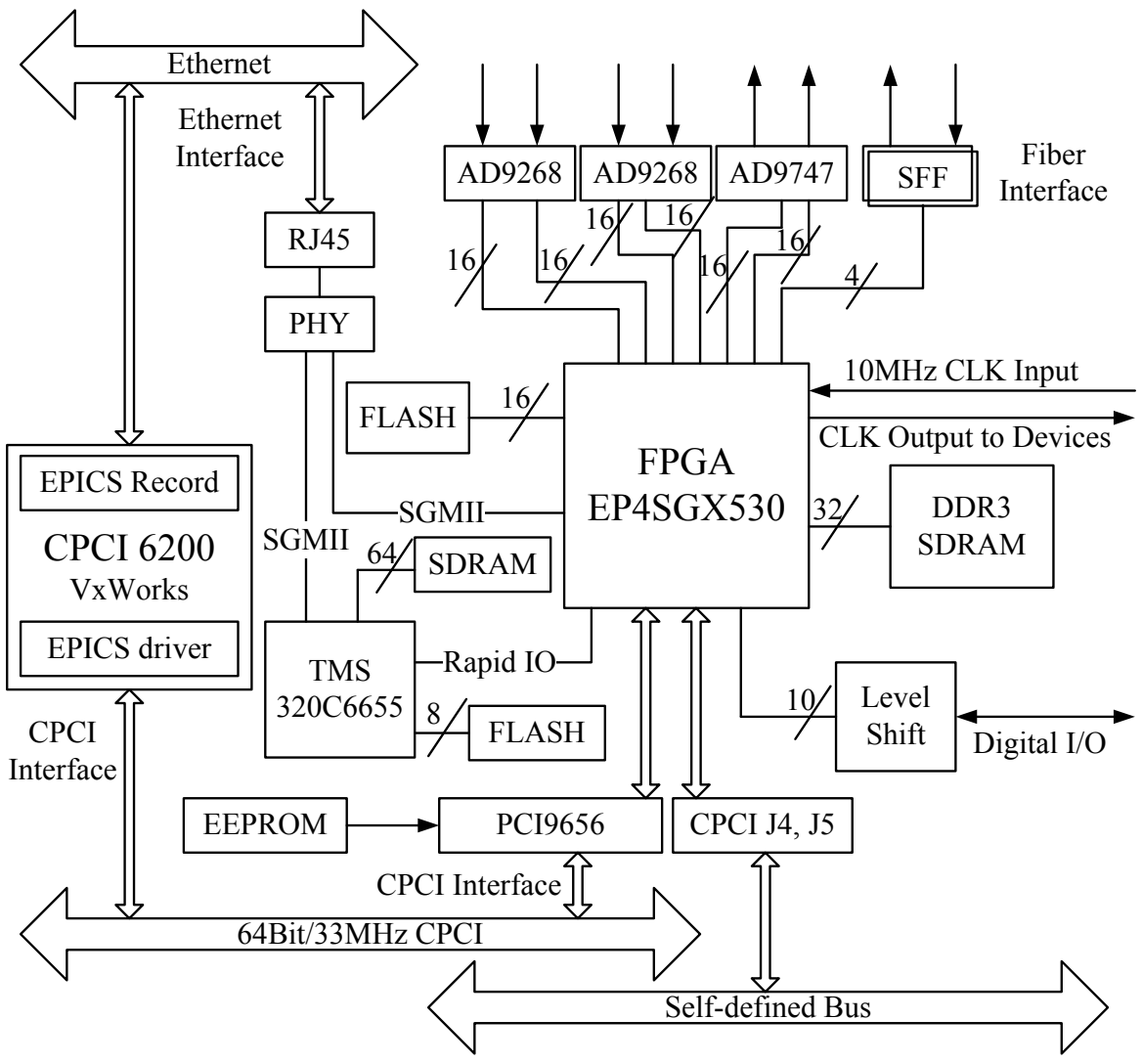

Fig. 4. LLRF system configuration.

## 4 Control architecture of LLRF system

Each RF system has independent LLRF system based on FPGA which is composed of 8 control loops including cavity voltage loop, cavity phase loop, synchronous phase loop, cavity tune loop, tetrode grid tune loop, beam loading compensation loop, orbit feedback loop, and direct RF feedback loop[5,6].Fig.5 shows the block diagram of the LLRF control system.

### 4.1 RF signal processing

For most control loops in CSNS/RCS LLRF control system, the procedure of signal processing works in much the same way. The RF and reference signal are both generated by DDS (Direct Digital Synthesizer) module. To satisfy the sweeping frequency work status of RF system, a direct demodulation method is adopted in LLRF system. All RF signals are sampled with 40MHz, and then digital signals are separately multiplied by two reference signals for orthogonal demodulation. A 70 taps FIR low-pass filter is adopted with 1.75μs delay. The Coordinate Rotation Digital Computer (CODIC) algorithm is also used to compute trigonometric function and get the amplitude and phase of the signals. Fig.6 shows the procedure of signal processing.

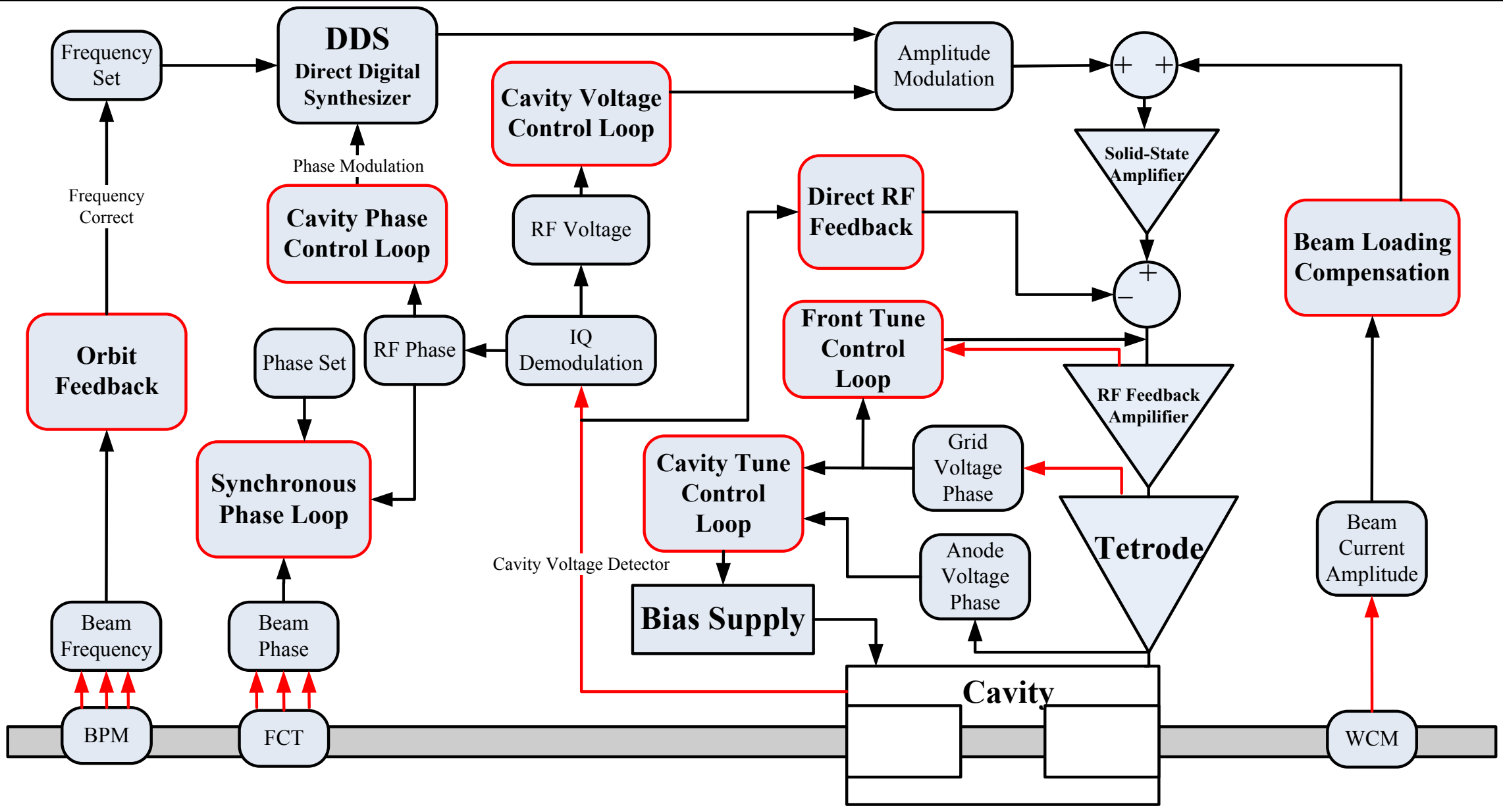

Fig. 5. Block diagram of the LLRF control system.

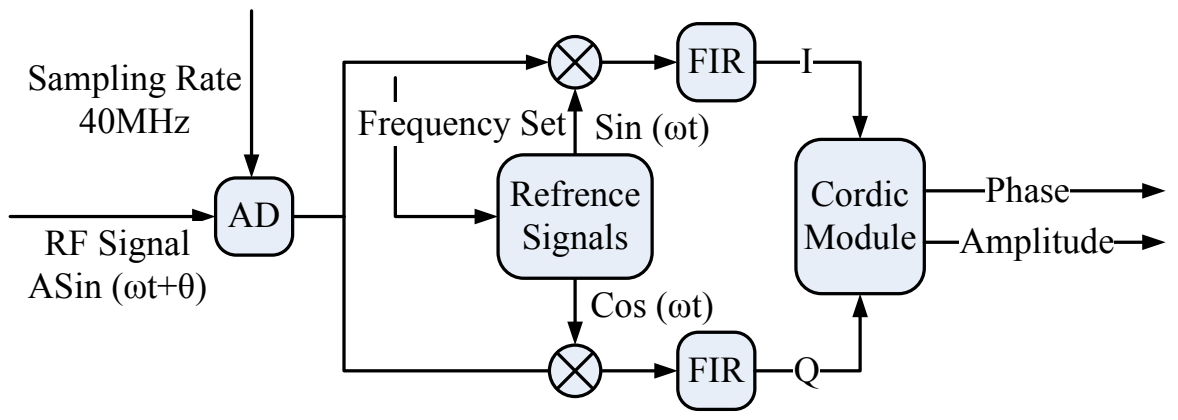

Fig. 6. Procedure of signal processing

## 4.2 Cavity Voltage Loop

The cavity voltage loop exists individually for each cavity at (h=2), and makes the cavity voltage follow a pattern shown in Fig.1. Each cavity has two acceleration gaps in parallel. For one gap, the minimum voltage is 1.3kV and the maximum voltage is 10.3kV.

There is a feedforward compensation option to decrease the following error induced by the rapid change of cavity voltage. The feedforward compensation mainly includes two parts, one is average of the amplitude modulation values of the past few cycles, and the other is the amplitude error multiplied by a certain factor.

## 4.3 Cavity phase Loop

The cavity phase loop locks the phase between accelerating voltage and RF reference signal. The initial phase of RF reference signal is triggered by the timing system, when the beam is injected into RCS. The cavity phase loop should compensate the phase shift mainly caused by 2 stages tune loops and frequency response in amplification chains. The bandwidth of cavity phase loop should be much higher than the tune loops to avoid loops coupling. When detuning happens, the cavity phase loop must immediately adjust the phase of RF driving signal to maintain the correct phase of accelerating voltage.

## 4.4 Synchronous Phase Loop

The synchronous phase loop damps synchrotron oscillation. The beam phase is detected by Fast Current Transformer (FCT), and compared with the phase of accelerating voltage. According to the result of comparison, the synchronous phase loop provides a damping component with 90 degree phase shift into the RF diver signal.

## 4.5 Cavity Tune Loop

The cavity tune loop is fed by the phase between grid voltage of tetrode and the cavity voltage. The change of resonant frequency of the cavity from 1.022MHz to 2.444MHz corresponds to the bias current changing from about 200A to 3000A.

Due to the bandwidth limitation of bias supply, a feedforward compensation is used for the 25Hz system operation. The algorithm is similar to cavity voltage loop. The actual bias current of the past few cycles and the tuning error are used to generate a feedforward correction table.

## 4.6 Tetrode Grid Tune Loop

The tetrode grid tune loop is used to compensate the parasitic capacitance of the tetrode grid, and ensure system gain and stability in operating frequency range. a linear bias supply of 10A is used to change the inductance of a ferrite-loaded low-Q resonant circuit, as shown in Fig.2. There is only a feedback loop for the tetrode grid tuning. Except for this, both of the tetrode grid tune loop and the cavity tune loop have the same operating principle.

### 4.7 Beam Loading Compensation Loop

The circulating current in CSNS/RCS is fairly high, and the beam loading effect must be carefully considered. A classic feedforward algorithm is adopted for the beam loading compensation. The beam signal is picked up by Wall Current Monitor (WCM), and added into the RF drive signal with an opposite phase. The fundamental component of the beam signal is taken out, and given the proper gain and phase[7].

Because of two stage of tuning loops, some phase error will be introduced. An adaptive algorithm is now being developed for adjusting the phase setting of the feedforward algorithm.

### 4.8 Orbit Feedback Loop

The orbit feedback loop is reserved to adjust the initial frequency setting. The beam signal is picked up by Beam Position Monitor (BPM), and used to compute the modification value.

### 4.9 Direct RF Feedback Loop

The RF feedback is an analog control loop. On account of the characteristics of low latency and high bandwidth, it is adopted to reduce transient beam loading and enhance the dynamic performance of the LLRF system. A small fraction of cavity voltage is picked up and fed back to the feedback amplifier. the phase shift in the feedback path is guaranteed by the two stage of tuning loops. Since there are no filters and signal demodulation, the RF feedback loop can response disturbance of arbitrary frequency. In practice, the feedback gain is about 20db.

## 5. System testing results

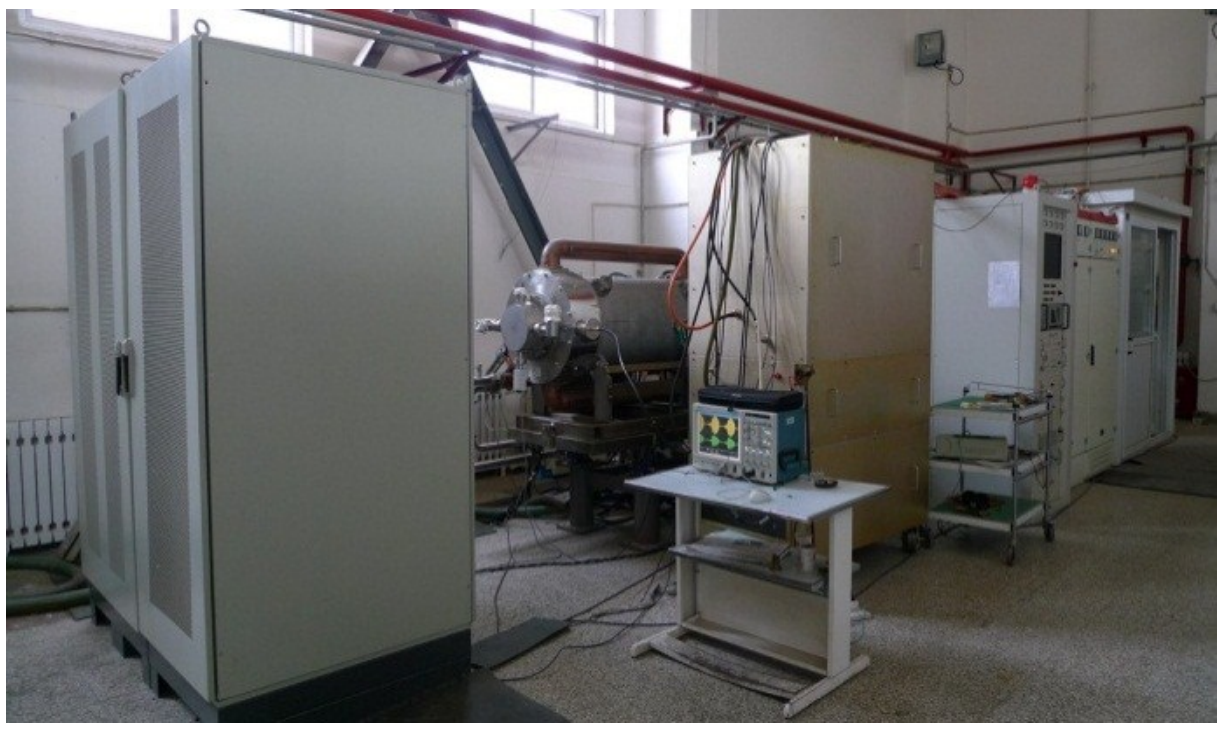

Fig. 7. View of the CSNS/RCS RF system prototypes.

The R&D prototype of RCS RF system have been developed, including a full size ferrite-loaded RF cavity, a set of RF amplifier, a 3300A bias supply and a digital LLRF system, as shown in Fig.7.

The high power integration test was completed on the prototype of RF system, with the cavity voltage loop, cavity phase loop, cavity tuning loop, tetrode grid tuning loop, and RF feedback loop. Except for the beam related control loops, the entire LLRF system is verified by the integration test. A schematic diagram of the test is shown in Fig.6. The maximum cavity voltage up to 12kV is achieved, with the RF frequency sweeping from 1.022 to 2.444MHz and a repetition rate of 25Hz. The waveform of cavity voltage and grid voltage of the tetrode is shown in Fig.8.

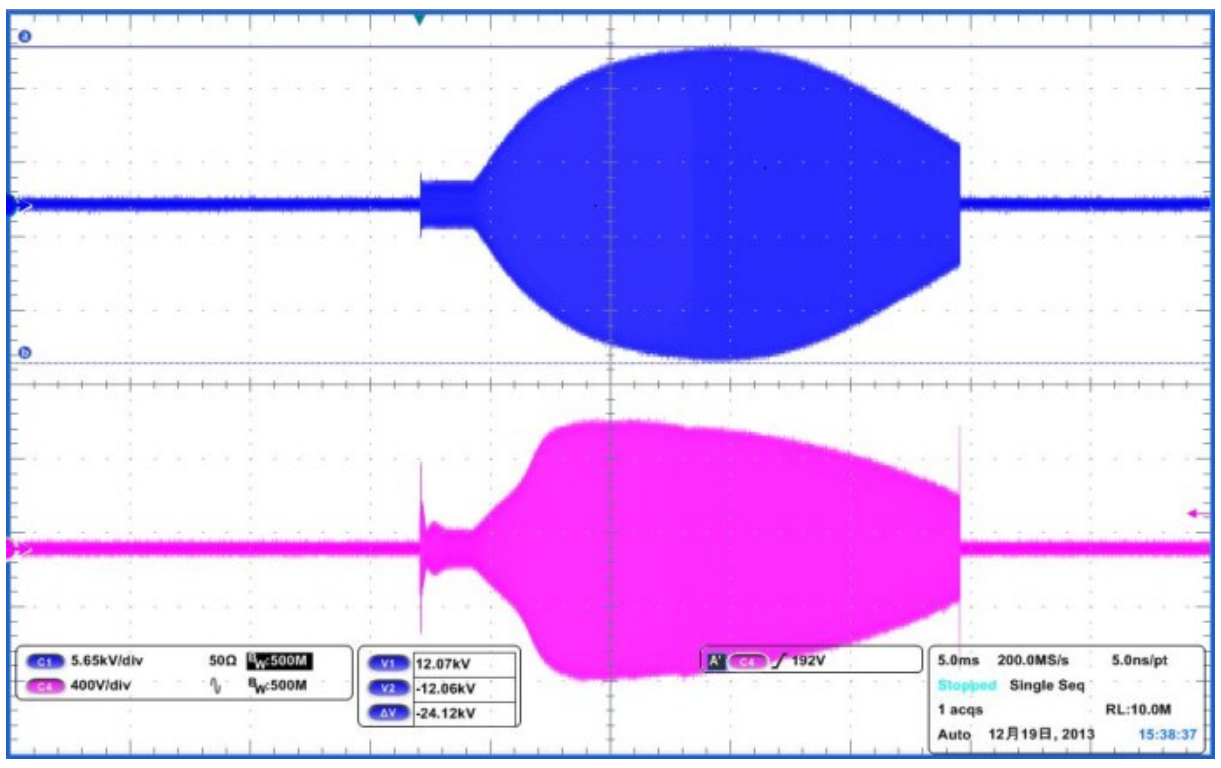

Fig. 8. Waveform of the cavity voltage (above) and grid voltage of the tetrode (below) on oscilloscope.

As shown in Fig.9, due to the dynamic range of cavity voltage, the amplitude error is more than 3% at beginning of one cycle with only feedback control, and decrease to less than ±0.4% with the feedforward compensation.

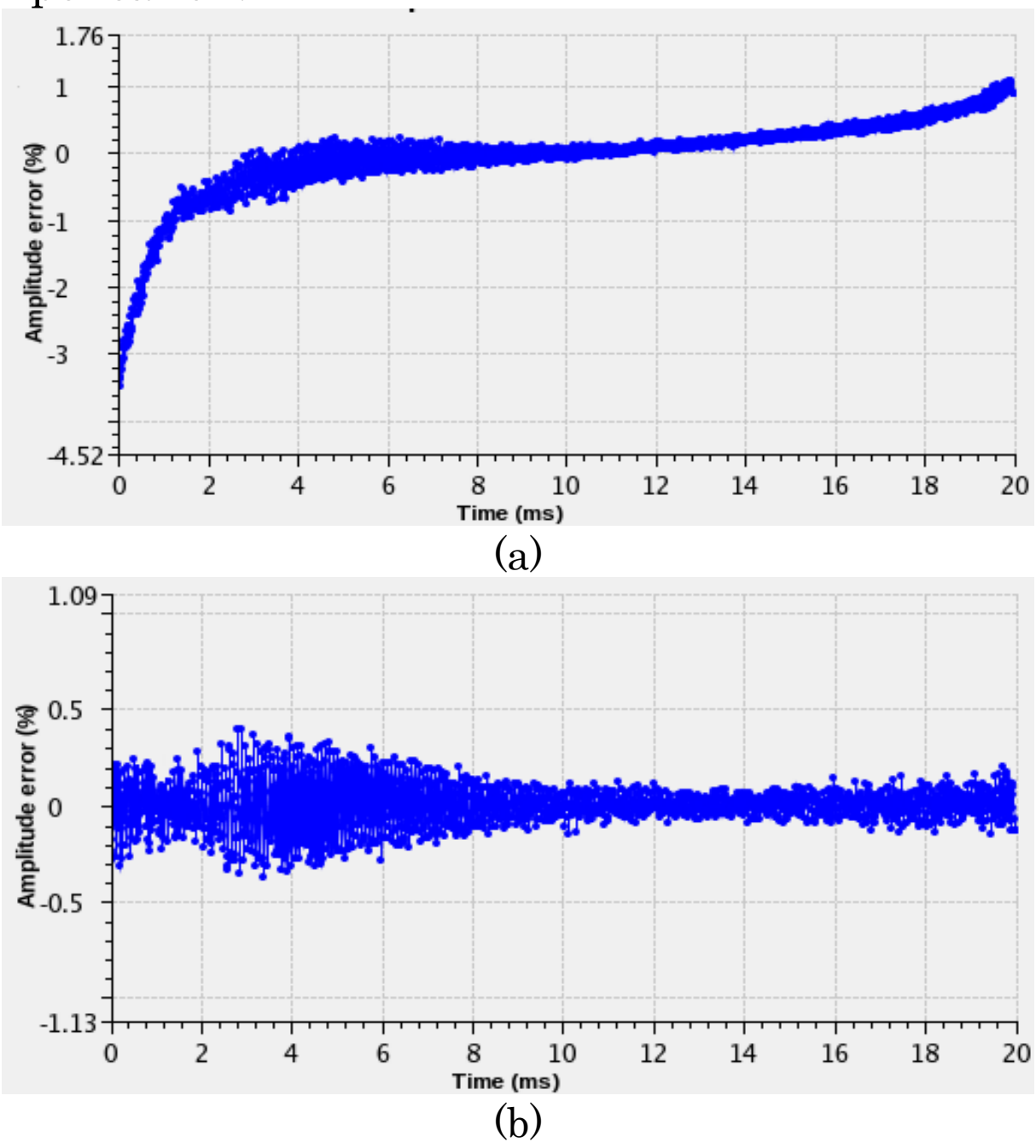

Fig. 9. Cavity voltage error in one cycle without (a) and with (b) feedforward compensation.

Fig.10 shows the phase error between the cavity voltage and the RF reference signal. With the cavity phase loop, the phase shift introduced by amplification chains including 2 stages tune loops, decrease to ±0.6 degree.

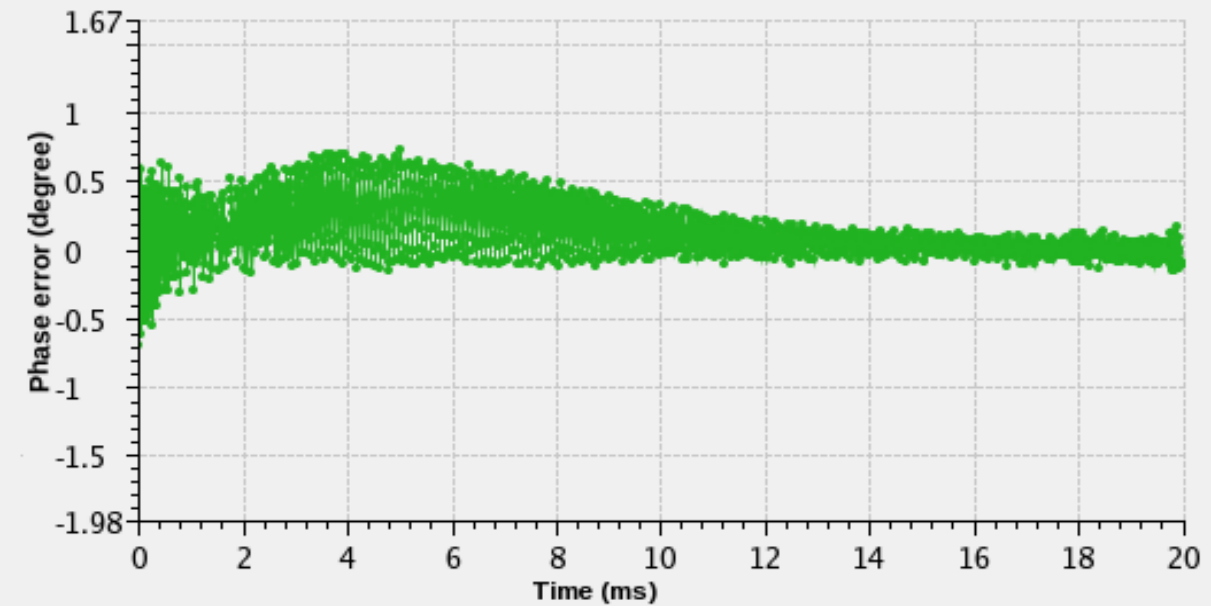

Fig. 10. Cavity phase error in one cycle.

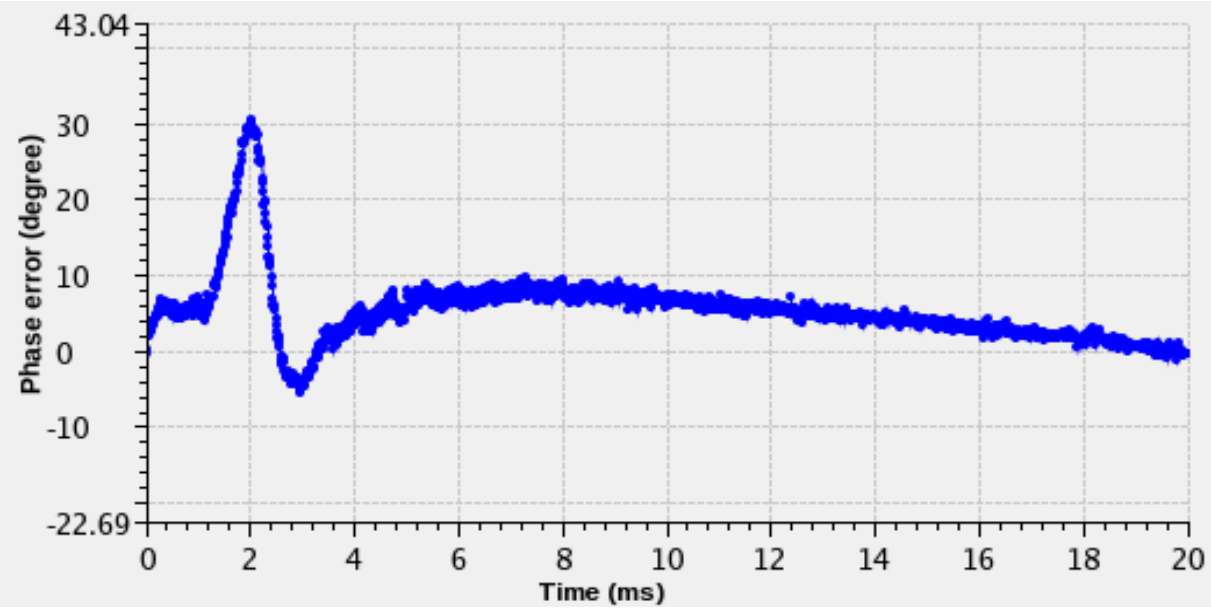

(a)

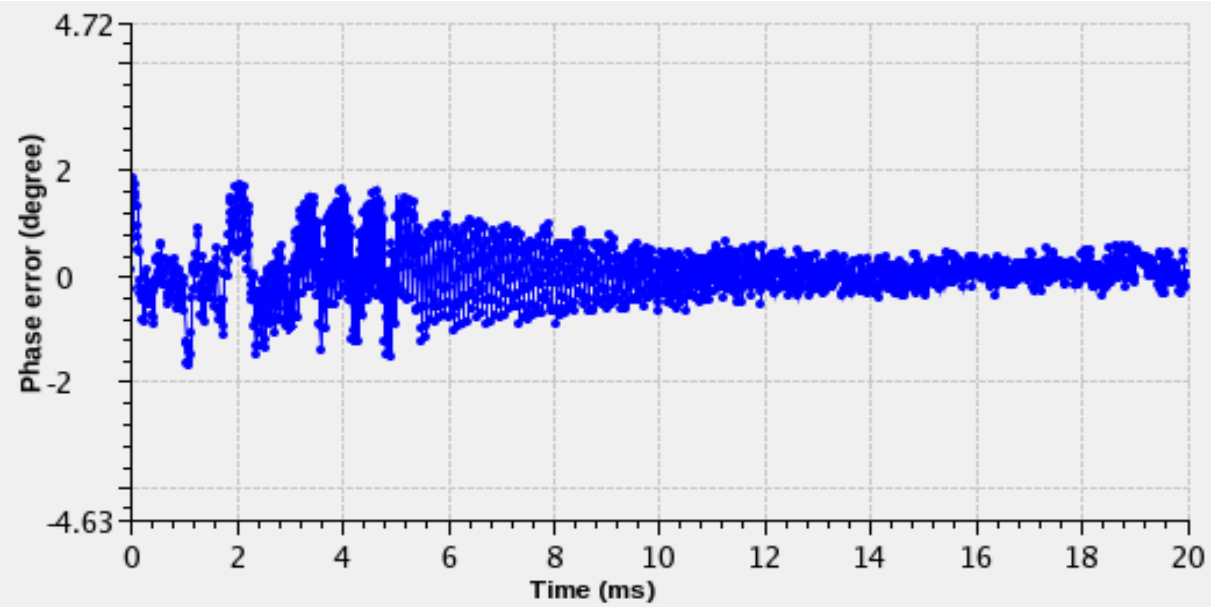

(b)

Fig. 11. Cavity tuning error in one cycle without (a) and with (b) feedforward compensation.

The algorithm of feedforward compensation used in cavity tune loop is similar to the cavity voltage loop. The maximum cavity tune error decreases from about 30 degree to 2 degree, as shown in Fig.11.

Because the Q value of the resonant circuit is much smaller than the cavity, the bandwidth of grid tune is higher than the cavity. Therefore, there is only a feedback control adopted. As shown in Fig.12, the maximum grid tune error is about 1.3 degree.

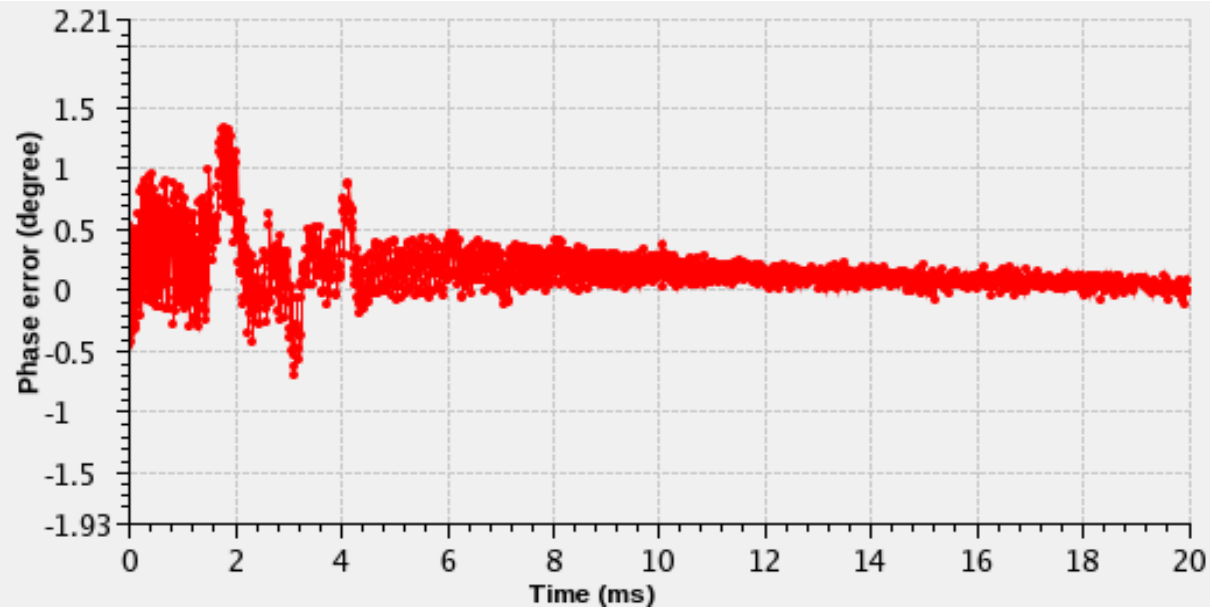

Fig. 12. Grid tuning error in one cycle.

## 5. Summary

The CSNS/RCS LLRF system provides a total solution for the RF system in RCS with rapid frequency sweeping and heavy beam loading effect. The cavity tuning and beam loading compensation are the key issues, and need to be properly handled. Feedforward compensation is widely used to improve dynamic performance and stability of the system. The bandwidth of different control loops should be carefully set to avoid coupling oscillation.